\documentclass[11pt]{article}

\usepackage[T1]{fontenc}
\usepackage[sc]{mathpazo}
\usepackage{amsmath}
\usepackage{amssymb}
\usepackage{enumerate}
\usepackage{amsthm}
\usepackage{amsfonts,mathrsfs}
\usepackage[style]{fncychap}
\usepackage{graphicx} 
\usepackage{geometry} 
\usepackage{eepic}
\usepackage{ifthen}
\newboolean{ElectronicVersion}
\setboolean{ElectronicVersion}{true} 

\usepackage{hyperref}
\hypersetup{pdfpagemode=UseNone}

\newcommand{\setft}[1]{\mathrm{#1}}
\newcommand{\lin}[1]{\setft{L}\left(#1\right)}

\def\be{\begin{equation}}
\def\ee{\end{equation}}
\def\bea{\begin{eqnarray*}}
\def\eea{\end{eqnarray*}}

\def\I{\mathbb{1}}

\newenvironment{mylist}[1]{\begin{list}{}{
    \setlength{\leftmargin}{#1}
    \setlength{\rightmargin}{0mm}
    \setlength{\labelsep}{2mm}
    \setlength{\labelwidth}{8mm}
    \setlength{\itemsep}{0mm}}}
    {\end{list}}

\def\ot{\otimes}

\newcommand{\iinner}[2]{\langle #1 | #2\rangle}
\newcommand{\out}[2]{| #1\rangle\langle #2 |}

\newcommand{\pa}[1]{(#1)}

\newcommand{\bra}[1]{\langle#1|}

\newcommand{\ket}[1]{|#1\rangle}

\DeclareMathOperator{\trace}{Tr}
\newcommand{\ptr}[2]{\trace_{#1}\pa{#2}}

\newcommand{\tr}[1]{\ptr{}{#1}}

\def\cD{\mathcal{D}}
\def\cH{\mathcal{H}}\def\cJ{\mathcal{J}}
\def\cK{\mathcal{K}}

\def\cV{\mathcal{V}}

\def\D{\textsf{D}}

\newtheorem{thrm}{Theorem}[section]

\theoremstyle{definition}

\numberwithin{equation}{section}

\newcounter{questionnumber}

\begin{document}

\title{\Large The Decomposition Properties of Quantum Discord}

\author{Sunho Kim\\{\small \it Department of Mathematics, Zhejiang
University, Hangzhou 310027, PR~China}\\Junde Wu\\{\small \it
Department of Mathematics, Zhejiang University, Hangzhou 310027,
PR~China}}
\date{}
\maketitle \mbox{}\hrule\mbox\\
\begin{abstract} The quantum discord was introduced by Ollivier, Zurek, Henderson and Vedral  as an indicator of the degree of
quantumness of mixed states. In this paper, we give out the
decomposition condition of quantum discords. Moreover, we show that
under the condition, the quantum correlations between the quantum
systems can be captured completely by the entanglement measure.
\end{abstract}
\mbox{}\hrule\mbox\\



{\bf $^*$ The corresponding author}:  Junde Wu, Department of
Mathematics, Zhejiang University, Hangzhou 310027, P. R. China, E-mail: wjd@zju.edu.cn

\section{Introduction and preliminaries}
In this article, we always assume that $\cH_A$,  $\cH_B$, $\cK_A$
and $\cK_B$ are finite dimensional complex Hilbert spaces. Let
$\lin{\cH_A, \cK_A}$ be the set of all linear operators from $\cH_A$
to $\cK_A$. A quantum state $\rho$ of some quantum system, described
by $\cH_A$, is a positive semi-definite operator of trace one, in
particular, for each unit vector $\ket{\psi} \in \cH_A$, the
operator $\rho = \out{\psi}{\psi}$ is said to be a \emph{pure
state}. We can identify the pure state  $\out{\psi}{\psi}$ with the
unit vector $\ket{\psi}$. The set of all quantum states on $\cH_A$
is denoted by $\D(\cH_A)$.

A quantum measurement is a set $\{M_{x}\}_{x\in\Sigma}$ of positive operators indexed by some
classical label $x$ corresponding to the classical outcomes of the
measurement. These operators form a
resolution of the identity on the Hilbert space of the system that
is being measured \cite{Davies, Holevo, Kraus}:
\begin{eqnarray*}
\forall x : M_{x} \geq 0,\ \ \sum_{x} M_{x} = \I,
\end{eqnarray*}
together with $\{A_{x}\}_{x\in\Sigma}$ such that $M_x = A_{x}^{\dag}A_{x}$. In particularly, when $\{M_x=\pi_{x}\}$ is a set of orthogonal
projection operators, then $\{M_x=\pi_{x}\}$ is said to be a von Neumann measurement.

Given a quantum state $\rho\in D(\cH_A)$, the quantum measurement $\{M_x\}$ induces a probability distribution $p=\{p_x\}_{x\in\Sigma}$, and the
conditional state $\rho_{A|x}$ given outcome $x$ and the probability
of this outcome read:
\begin{eqnarray*}
\rho_{A|x} = p(x)^{-1}A_{x}\rho A_{x}^{\dag}, \ \ \ \ p(x) = \tr{M_{x}\rho}.
\end{eqnarray*}

However, the following famous theorem told us that each quantum measurement can be seen as a von Neumann measurement on a larger quantum system, that is:

\begin{thrm}\label{th:Neumark theorem}(\textbf{Neumark extension theorem}, \cite{Spehner, Watrous})
Let $M=\{M_x\}_{x\in\Sigma}$ be a quantum measurement on $\cH_A$
with $|\Sigma| = n$. Then there exist a Hilbert space $\cH_E$ with
dimension $\dim{\cH_E} = n$, a pure state $\ket{\epsilon_0}\in
\cH_E$, a von Neumann measurement $\{\pi^E_x\}$ on $\cH_E$, and a
unitary operator $U$ on $\cH_{A}\ot\cH_{E}$ such that for each
quantum state $\rho\in\D(\cH_A)$, \bea A_x\rho A_x^{\dag} =
\ptr{E}{\I_A \ot \pi^E_x\ U\ \rho \ot \out{\epsilon_0}{\epsilon_0}\
U^{\dag}\ \I_A \ot \pi^E_x}, \eea where $M_x = A_{x}^{\dag}A_{x}$.
\end{thrm}
It follows from the theorem that (\cite{Spehner})
\bea
M_x = A_{x}^{\dag}A_{x} = \bra{\epsilon_0}U^{\dag}\I_A \ot \pi^E_xU\ket{\epsilon_0},
\eea
and the probability of the outcome $x$ read
\bea
p_x = \tr{M_{x}\rho} = \tr{U^{\dag}\I_A \ot \pi^E_xU\ \rho \ot \out{\epsilon_0}{\epsilon_0}}.
\eea

Let $p = \{p_{a}\}\in \mathbb{R}^{\Sigma}$ be a probability
distribution, the \emph{Shannon entropy} $H(p)$ of $p$  is defined
by \cite{Shannon}
\begin{eqnarray*}
H(p) = -\sum_{a\in\Sigma}p_{a}\log_2(p_{a}).
\end{eqnarray*}

For each quantum state $\rho\in\D(\cH_A)$, the quantum analog of the
Shannon entropy is \emph{von Neumann entropy}
\begin{eqnarray*}
S(\rho) = - \tr{\rho\log_2(\rho)}.
\end{eqnarray*}

An equivalent expression of von Neumann entropy is (\cite{Spehner}, Chapter 6.1)
\begin{eqnarray*}
S(\rho) = \min_{\{\ket{\psi_{i}}, p_{i}\}} H(\{p_{i}\}),
\end{eqnarray*}
where the minimum is over all pure state convex decompositions of $\rho$.

Moreover, a pure state convex decomposition $\{\ket{\psi^{0}_{i}},
p^{0}_{i}\}$ of $\rho$ minimizes $\{H(\{p_{i}\}): \{\ket{\psi_{i}},
p_{i}\}\}$ if and only if it is a spectral decomposition of $\rho$.

The identity can be generalized to get \be\label{eq:entropy} S(\rho)
\leq H(\{\eta_{i}\}) + \sum_{i}\eta_{i}S(\rho_{i}) \ee for any
quantum state ensemble $\{\rho_{i}, \eta_{i}\}$, where $\{\rho_{i},
\eta_{i}\}$ is a convex decomposition of $\rho$. Moreover, it has
equality if and only if the quantum states $\{\rho_{i}\}$ have
mutual orthogonal supports.

Let us consider two quantum systems $\cH_A$ and $\cH_B$, $\rho_{AB}\in D(\cH_A\ot\cH_B)$.
In quantum information theory, the quantum mutual information
\begin{eqnarray*}
I_{A:B}(\rho_{AB}) = S(\rho_{A}) + S(\rho_{B}) - S(\rho_{AB})
\end{eqnarray*} of the quantum state $\rho_{AB}$ is regarded as a measure of the total correlations between quantum systems $\cH_A$ and $\cH_B$ when the quantum system  $\cH_A\ot\cH_B$ in the quantum state $\rho_{AB}$, where $\rho_{A} = \ptr{B}{\rho_{AB}}$ and $\rho_{B} = \ptr{A}{\rho_{AB}}$ are the reduced states of $\rho_{AB}$.

If we denote $S(\rho_{B|A}) = S(\rho_{AB}) - S(\rho_{A})$, then the quantum mutual information can be written in the following form:
\begin{eqnarray*}
 I_{A:B}(\rho_{AB}) = S(\rho_{B}) - S(\rho_{B|A}).
\end{eqnarray*}

Ones can prove that $I_{A:B}(\rho_{AB}) \geq 0$ and
$I_{A:B}(\rho) = 0$ if and only if $\rho_{AB}$ is a product state, that
is $\rho_{AB} = \rho_{A}\ot\rho_{B}$.

Given a von Neumann measurement $\{\pi^{A}_{x}\}$ on quantum system $\cH_A$, let us defined a conditional entropy on quantum system $\cH_B$
by $$S_{B|A}(\rho_{AB}|\{\pi^{A}_{x}\}) = \sum_{i}\eta_{x}S(\rho_{B|x}),$$
where
\begin{eqnarray*}
\rho_{B|x} = \eta^{-1}_{x}\ptr{A}{\pi^{A}_{x}\otimes \I_{B}\rho_{AB}}, \
\  \ \ \eta_{x} = \tr{\pi^{A}_{x}\otimes \I_{B}\rho_{AB}}.
\end{eqnarray*}
Denote
\begin{eqnarray*}
\cJ\{\pi^{A}_{x}\}(\rho_{AB}) = S(\rho_{B}) -
S_{B|A}(\rho_{AB}|\{\pi^{A}_{x}\}).
\end{eqnarray*}
In order to take a quantity which does not depend on the von Neumann measurements, ones define
\begin{eqnarray*}
\cJ_{B|A}^{v.N.}(\rho_{AB}) = \max_{\{\pi^{A}_{x}\}}
\cJ\{\pi^{A}_{x}\}(\rho_{AB}) = S(\rho_{B}) -
\min_{\{\pi^{A}_{x}\}}\{\sum_{x}\eta_{x}S(\rho_{B|x})\},
\end{eqnarray*}
it is interpreted as a measure of classical correlations between the quantum systems $\cH_A$ and $\cH_B$ when the quantum system  $\cH_A\ot\cH_B$ in the quantum state $\rho_{AB}$.

In general, $I_{A:B}(\rho_{AB})$ may differ $\cJ_{B|A}^{v.N.}(\rho_{AB})$. Their difference
\begin{eqnarray*}
\cD_{A}^{v.N.}(\rho_{AB}) = I_{A:B}(\rho_{AB}) - \cJ_{B|A}^{v.N.}(\rho_{AB}) =
S(\rho_{A}) - S(\rho_{AB}) +
\min_{\{\pi^{A}_{x}\}}\{\sum_{x}\eta_{x}S(\rho_{B|x})\}
\end{eqnarray*}
is interpreted as a measure of quantum correlations and is called
quantum discord (\cite{Spehner}, Chapter 10.1 and \cite{Ollivier,
Henderson, Vedral}). The minimum is achieved for some rank-one
orthogonal projection measurement operators $\{\pi^{A}_{x}\}$.

Similarly, given a quantum measurement $\{M^{A}_{z}\}$ on $\cH_A$, let
us defined a conditional entropy on $\cH_B$ by
$$S_{B|A}(\rho_{AB}|\{M^{A}_{z}\}) = \sum_{z}\mu_{z}S(\rho_{B|z}),$$
where
\begin{eqnarray*}
\rho_{B|z} = \mu^{-1}_{z}\ptr{A}{M^{A}_{z}\otimes \I_{B}\rho_{AB}} \ \ ,
\ \ \mu_{z} = \tr{M^{A}_{z}\otimes \I_{B}\rho_{AB}}.
\end{eqnarray*}
Denote
\begin{eqnarray*}
\cJ\{M^{A}_{z}\}(\rho_{AB}) = S(\rho_{B}) - S_{B|A}(\rho_{AB}|\{M^{A}_{z}\}),
\end{eqnarray*} and
\begin{eqnarray*}
\cJ_{B|A}(\rho_{AB}) = \max_{\{M^{A}_{z}\}} \cJ\{M^{A}_{z}\}(\rho_{AB}) =
S(\rho_{B}) - \min_{\{M^{A}_{z}\}}\{\sum_{z}\mu_{z}S(\rho_{B|z})\}.
\end{eqnarray*}
The corresponding discord $\cD_{A}(\rho_{AB})$ is defined by (\cite{Spehner})
\begin{eqnarray*} \cD_{A}(\rho_{AB}) = I_{A:B}(\rho_{AB}) - \cJ_{B|A}(\rho_{AB}) =
S(\rho_{A}) - S(\rho_{AB}) +
\min_{\{M^{A}_{z}\}}\{\sum_{z}\mu_{z}S(\rho_{B|z})\}.
\end{eqnarray*}

As in the case of von Neumann measurements, the minimum is achieved
for some rank-one measurement operators $\{M^{A}_{z}\}$.

That $\cD_{A}(\rho_{AB})
\leq \cD_{A}^{v.N.}(\rho_{AB})$ is clear. On the other hand, by Neumark extension
theorem (\ref{th:Neumark theorem}) and note that $M^{A}_{z} =
\bra{\epsilon_0}U^{\dag}\I_A \ot \pi^E_zU\ket{\epsilon_0},$ we have
\begin{eqnarray}\label{discord-1}
\cD_{A}(\rho_{AB}) = \cD_{AE}^{v.N.}(\rho_{AB} \otimes
\out{\epsilon_{0}}{\epsilon_{0}}).
\end{eqnarray}

Given a pure state $\out{\psi}{\psi}_{AB}\in \D(\cH_{A}\ot\cH_{B})$,
then $S(\rho_{A})=S(\rho_{B})$ \cite{Bennett}. The entanglement
$E(\out{\psi}{\psi}_{AB})$ of $\out{\psi}{\psi}_{AB}$ is defined by
$$E(\out{\psi}{\psi}_{AB}) = S(\rho_{A}) = S(\rho_{B}).$$

For any quantum state $\rho_{AB}\in D(\cH_{A}\ot\cH_{B})$, the
\emph{entanglement of formation $E_{f}(\rho_{AB})$} of $\rho_{AB}$
is defined by (\cite{{Bennett}}, \cite{Wootters}):
\begin{eqnarray*}
E_{f}(\rho_{AB}) = \min_{\{\ket{\psi_{i}}, p_{i}\}}
\sum_{i}p_{i}E(\out{\psi_{i}}{\psi_{i}}),
\end{eqnarray*}
where $\{\out{\psi_{i}}{\psi_{i}}, p_i\}_{i\in\Sigma}$ is the pure
state convex decomposition of $\rho_{AB}$.

For any pure state $\out{\psi}{\psi}_{AB}\in D(\cH_A\ot\cH_B)$, we
have (\cite{Bennett, Wootters})
\begin{eqnarray}\label{discord-2}
\cD_{A}(\out{\psi}{\psi}_{AB}) = \cD_{A}^{v.N.}(\out{\psi}{\psi}_{AB}) = E_{f}(\out{\psi}{\psi}_{AB}) = S(\out{\psi}{\psi}_{A}) = S(\out{\psi}{\psi}_{B}).
\end{eqnarray}

Let $\rho_{AB}\in\D(\cH_A\otimes\cH_B)$, $\rho_{AB}
=\sum_{i\in\Sigma}p_{i}\out{u_i}{u_i}$ be its spectral
decomposition. Generally, ones have (\cite{Spehner}) \bea 0 \leq
\min_{\{M^A_z\}}\{\sum_z\mu_zS(\rho_{B|z})\} \leq
\min_{\{\pi^A_x\}}\{\sum_x\eta_xS(\rho_{B|x})\} \leq S(\rho_{AB}).
\eea Moreover, \bea \min_{\{\pi^A_x\}}\{\sum_x\eta_xS(\rho_{B|x})\}
= S(\rho_{AB}) \eea if and only if
$\sum_{i\in\Sigma}p_{i|y}\out{\phi_{iy}}{\phi_{iy}}$ is the spectral
decomposition of $\rho_{B|y}$, where
$$p_{i|y} = \frac{p_i\tr{\pi^A_y\ot\I_B\out{u_i}{u_i}}}{\eta_y}, \quad \ \quad \out{\phi_{iy}}{\phi_{iy}} = \frac{\ptr{A}{\pi^A_y\ot\I_B\out{u_i}{u_i}}}{\tr{\pi_y\ot\I_B\out{u_i}{u_i}}},$$
the von Neumann measurement $\{\pi^A_y\}$ minimizes the conditional
entropy $\sum_x\eta_xS(\rho_{B|x})$, and $\ptr{B}{\out{u_i}{u_j}}=0$ when $i\neq j$ and $p_ip_j>0$.

In this paper, we give out the decomposition condition of quantum
discords. Moreover, we show that under the condition, the quantum
correlations between the quantum systems can be captured completely
by the entanglement measure.

\section{Decomposition of quantum discord}

Firstly, we prove the following result.

\begin{thrm}\label{th:thrm-1} Let $\rho_{AB}\in\D(\cH_A\otimes\cH_B)$, $\rho_{AB}=\sum_{i\in\Sigma}p_{i}\out{u_i}{u_i}$ be its spectral decomposition. Then \be\label{eq:condition-2}
\min_{\{\pi^A_x\}}\{\sum_{x}\eta_{x}S(\rho_{B|x})\} = 0
\ee
if and only if for any two $i,j\in\Sigma$ and $i\neq j$, \be\label{eq:condition-1}
\ptr{A}{\out{u_i}{u_j}} = 0.
\ee
\end{thrm}

\begin{proof} If for any two $i,j\in\Sigma$ and $i\neq j$, we have $\ptr{A}{\out{u_i}{u_j}} = 0$, then
$\ptr{B}{\out{u_i}{u_i}}$ and $\ptr{B}{\out{u_j}{u_j}}$ are orthogonal, it implies that there are subspaces $\cV_{i}^{A}, \cV_{j}^{A}\subseteq \cH_{A}$ such that $\out{u_i}{u_i} \in \D(\cV_{i}^{A}\otimes \cH_{B})$ and $\cV_{i}^{A}\subseteq (\cV_{j}^{A})^{\perp}$. Let $\pi^{A}_{i}$ be the orthogonal projector onto $\cV_{i}^{A}$ for any $i$, and $\pi^{A}_{m} = \I_{A} - \{\sum_{i\in\Sigma}\pi^{A}_{i}\}$, then
\bea
\pi^{A}_{i}\otimes \I_{B}\ \rho_{AB} \ \pi^{A}_{i}\otimes \I_{B} = p_i\out{u_i}{u_i}.
\eea
Therefore, we have that
\bea
0\leq \min_{\{\pi^A_x\}}\{\sum_x \eta_x S(\rho_{B|x})\} \leq \sum_{i\in\Sigma\cup\{m\}}p_i \min_{\{\pi_k^{(i)}\}}\{\sum_{k\in\Sigma_i} \eta^{(i)}_k S(\rho_{B|k}^{(i)})\},
\eea
where $\sum_{k\in\Sigma_i} \pi_k^{(i)} = \pi^A_i,\ \eta^{(i)}_k = \tr{\pi^{(i)}_{k}\otimes \I_{B}\out{u_i}{u_i}}$ and $\rho_{B|k}^{(i)} = (\eta^{(i)}_k)^{-1}\ptr{A}{\pi^{(i)}_{k}\otimes \I_{B}\out{u_i}{u_i}}$, $\{\pi_k^{(i)}\}_{k\in\Sigma_i,i\in\Sigma\cup\{m\}}$ is also a von Neumann measurement. On the other hand, for any pure state, it follows from the Schmidt decomposition that states $\{\rho_{B|k}^{(i)}\}$ are pure states and thus have zero entropy. Therefore, we have that
\bea
\min_{\{\pi^A_x\}}\{\sum_x \eta_x S(\rho_{B|x})\} = 0.
\eea

If there exist $i\neq j$ and $p_i,p_j>0$ such that $\ptr{A}{\out{u_i}{u_j}}\neq 0$, then for any von Neumann measurement $\{\pi^A_x\}_x$ on $\cH_A$, there exists at least a $\pi^A_x$ such that $\ptr{A}{\pi^A_x\otimes \I_A\out{u_k}{u_k}}\neq 0$ for $k=i,j$. Also, note that  $\sum_{i\in\Sigma}p_{i}\out{u_i}{u_i}$ is the spectral decomposition of $\rho$, we have that $\langle v^{(i)}_x|v^{(j)}_x\rangle = 0$, where $\ptr{A}{\pi_x\otimes \I_A\out{u_k}{u_k}} = \eta^{(k)}_x\out{v^{(k)}_x}{v^{(k)}_x}$ for $k=i,j$. It is easy to show that $S(\rho_{B|x}) > 0.$ This  contradicts (\ref{eq:condition-2}).
\end{proof}

\begin{thrm}\label{th:thrm-2}
Let $\rho_{AB}\in\D(\cH_A\otimes\cH_B)$, $\rho_{AB} = \sum_{i\in\Sigma}p_{i}\out{u_i}{u_i}$ be its spectral decomposition. If
\bea
\ptr{A}{\out{u_i}{u_j}} = 0
\eea
for any two $i\neq j$ and $p_i,p_j>0$, then
\be
\cD_{A}(\rho_{AB}) = \cD_{A}^{v.N.}(\rho_{AB}) = E_f(\rho_{AB}) = \sum_{i\in\Sigma}p_{i}\cD_{A}^{v.N.}(\out{u_i}{u_i}).
\ee
\end{thrm}

\begin{proof} If $\ptr{A}{\out{u_i}{u_j}} = 0$ for any two $i\neq j$ and $p_i, p_j>0$, note that the property of entropy, we know that
\bea
S(\rho_A) = \sum_{i\in\Sigma} p_i S(\rho_{A|i}) + H(\{p_i\}),
\eea
where $\rho_{A|i} = \ptr{B}{\out{u_i}{u_i}}.$
And, it follows from  the spectral decomposition
$\rho_{AB} = \sum_{i\in\Sigma}p_{i}\out{u_i}{u_i}$ that $S(\rho_{AB}) = H(\{p_i\}).$
Therefore, by the definition of quantum discord, equality (\ref{discord-2}) and Theorem \ref{th:thrm-1}, we have that
\bea
\cD_{A}^{v.N.}(\rho_{AB}) &=& S(\rho_{A}) - S(\rho_{AB}) + \min_{\{\pi^A_x\}}\{\sum_{x}\eta_{x}S(\rho_{B|x})\}\\
&=& \sum_{i\in\Sigma} p_i S(\rho_{A|i}) + H(\{p_i\}) - H(\{p_i\})\\
&=& \sum_{i\in\Sigma}p_{i}\cD_{A}^{v.N.}(\out{u_i}{u_i}).
\eea

Also, by Equality (\ref{discord-1}) and (\ref{discord-2}), we have that
\begin{eqnarray*}
\cD_{A}(\rho_{AB})&=&\cD_{AE}^{v.N.}(\rho_{AB} \otimes
\out{\epsilon_{0}}{\epsilon_{0}}) = \sum_{i\in\Sigma}p_{i}\cD_{AE}^{v.N.}(\out{u_i}{u_i} \otimes
\out{\epsilon_{0}}{\epsilon_{0}})\\
&=&\sum_{i\in\Sigma}p_{i}\cD_{A}^{v.N.}(\out{u_i}{u_i}) = \cD_{A}^{v.N.}(\rho_{AB}).
\end{eqnarray*}

Moreover, it follows from  \cite{Horodecki} that
\begin{eqnarray*}
E_{f}(\rho_{AB}) = \sum_{i}p_{i}E_{f}(\out{u_i}{u_i}),
\end{eqnarray*}
thus,  by Equality (\ref{discord-2}) again, we have \be
\cD_{A}(\rho_{AB}) = \cD_{A}^{v.N.}(\rho_{AB}) = E_f(\rho_{AB} =
\sum_{i\in\Sigma}p_{i}\cD_{A}^{v.N.}(\out{u_i}{u_i}). \ee
\end{proof}

This theorem means that if $\sum_{i\in\Sigma}p_{i}\out{u_i}{u_i}$ is
a spectral decomposition of $\rho_{AB}$ and \bea
\ptr{A}{\out{u_i}{u_j}} = 0 \quad \textrm{where}\  i\neq j\
\textrm{and}\  p_i,p_j>0, \eea then the quantum correlations between
the quantum systems $\cH_A$ and $\cH_B$ can be captured completely
by the entanglement measure.

If we replace the condition of pure states in equality
(\ref{eq:condition-1}) with mixed states, we have the following
conclusion:

\begin{thrm}\label{th:thrm-3}
Let $\rho_{AB}\in\D(\cH_A\otimes\cH_B)$, $\rho_{AB} = \sum_{i\in\Sigma}p_{i}\rho_{i}$ be its an orthogonal decomposition. If for
any two $i\neq j\in\Sigma$ and $p_i,p_j>0$,
\begin{eqnarray}
\ptr{A}{\rho_{i}\rho_{j}} = 0,
\end{eqnarray}
then
\be
\cD_{A}(\rho_{AB}) = \sum_{i\in\Sigma}p_{i}\cD_{A}(\rho_{i}).
\ee
\end{thrm}

\begin{proof} If $\ptr{A}{\rho_i\rho_j} = 0$ for any two $i,j\in\Sigma$ and $i\neq j$, then it is easy to show that
$\ptr{B}{\rho_i}$ and $\ptr{B}{\rho_j}$ are orthogonal, it implies that there are subspaces $\cV_{i}^{A}, \cV_{j}^{A} \subseteq \cH_{A}$ such that $\rho_i \in \D(\cV_{i}^{A}\otimes \cH_{B})$ and $\cV_{i}^{A}\subseteq (\cV_{j}^{A})^{\perp}$. If $\pi^{A}_{i}$ is the orthogonal projector onto $\cV_{i}^{A}$ for any $i$, and $\pi^{A}_{m} = \I_{A} - \{\sum_{i\in\Sigma}\pi^{A}_{i}\}$, we have
\bea
\pi^{A}_{i}\otimes \I_{B}\ \rho_{AB} \ \pi^{A}_{i}\otimes 1_{B} = p_i\rho_i.
\eea

Let $\{M_z = A_z^{\dag}A_z\}_{z}$ be the quantum measurement which minimizes the conditional entropy $\sum_z \mu_zS(\rho_{B|z})$. Note that
$A_z^{\dag}\pi^{A}_{i}A_z$ are positive operator for all $i,z$, and
\bea
\sum_{i,z}A_z^{\dag}\pi^{A}_{i}A_z = \sum_z A_z^{\dag}(\sum_i\pi^{A}_{i})A_z = \sum_{z}A_z^{\dag}A_z = \I_A,
\eea
thus, $\{M_z^{(i)} = A_z^{\dag}\pi^{A}_{i}A_z\}_{i\in\Sigma\cup\{m\},z}$ is also a quantum measurement. By
\bea
\ptr{A}{M_z^{(i)}\otimes \I_{B}\rho_{AB}} = p_i\ptr{A}{M_z^{(i)}\otimes \I_{B}\rho_{i}}
\eea
for all $i,z$, we have
\be\label{eq:1}
\sum_z \mu_z S(\rho_{B|z}) \leq \sum_{i\in\Sigma}p_i \{\sum_z \mu^{(i)}_z S(\rho_{B|z}^{(i)})\},
\ee
where $\mu^{(i)}_z = \tr{M_z^{(i)}\otimes \I_{B}\rho_i}$ and $\rho_{B|z}^{(i)} = (\mu^{(i)}_z)^{-1}\ptr{A}{M_z^{(i)}\otimes \I_{B}\rho_i}$.

On the other side, it follows from $\mu_z\rho_{B|z} = \sum_i p_i\mu^{(i)}_z\rho_{B|z}^{(i)}$ and the concavity of von Neumann entropy that
\be\label{eq:2}
\sum_z \mu_z S(\rho_{B|z}) \geq \sum_{i\in\Sigma,z}p_i\mu^{(i)}_z S(\rho_{B|z}^{(i)}) = \sum_{i\in\Sigma}p_i \{\sum_z \mu^{(i)}_z S(\rho_{B|z}^{(i)})\}.
\ee

Therefore, by Inequality (\ref{eq:1}) and (\ref{eq:2}), we have
\bea
\min_{\{M_z\}}\sum_z \mu_z S(\rho_{B|z}) = \sum_{i\in\Sigma}p_i \min_{\{M_z^{(i)}\}}\{\sum_z \mu^{(i)}_z S(\rho_{B|z}^{(i)})\}.
\eea

Moreover, if denote $\rho_{A|i} = \ptr{B}{\rho_i}$,  then, it follows from $\rho_A = \sum_{i\in\Sigma} p_i\rho_{A|i}$ and property of entropy (\ref{eq:entropy}) that
\bea
S(\rho_A) = \sum_{i\in\Sigma} p_i S(\rho_{A|i}) + H(\{p_i\}).
\eea
Similarly, we also have
\bea
S(\rho) = \sum_{i\in\Sigma} p_i S(\rho_{i}) + H(\{p_i\}).
\eea
Therefore, by the definition of quantum discord, it follows that
\bea
\cD_{A}(\rho_{AB}) &=& S(\rho_{A}) - S(\rho) + \min_{\{M_z\}}\{\sum_{z}\mu_{z}S(\rho_{B|z})\}\\
&=& \sum_{i\in\Sigma} p_i S(\rho_{A|i}) + H(\{p_i\}) - \{\sum_{i\in\Sigma} p_i S(\rho_{i}) + H(\{p_i\})\} + \sum_{i\in\Sigma}p_i \min_{\{M_z^{(i)}\}}\{\sum_z \mu^{(i)}_z S(\rho_{B|z}^{(i)})\}\\
&=& \sum_{i\in\Sigma}p_{i}\big[S(\rho_{A|i}) - S(\rho_{i}) + \min_{\{M_z^{(i)}\}}\{\sum_z \mu^{(i)}_z S(\rho_{B|z}^{(i)})\}\big]\\
&=& \sum_{i\in\Sigma}p_{i} \cD_{A}(\rho_i).
\eea
\end{proof}

\section{A tripartite system}

Let $\rho_{AB}\in\D(\cH_A\otimes\cH_B)$. Taking a pure state
$\out{\Psi}{\Psi}_{ABC}\in \D(\cH_{A}\ot\cH_{B}\ot\cH_{C})$ such
that \bea  \rho_{AB} = \ptr{C}{\out{\Psi}{\Psi}_{ABC}}, \eea
$\rho_{AB} = \sum_{i\in\Sigma}p_{i}\out{u_i}{u_i}$ is the spectral
decomposition of $\rho_{AB}$. Now, we will prove that if $\rho_{AB}$
satisfies the condition of Theorem \ref{th:thrm-2}, then by the
famous necessary and sufficient condition of zero discord in
\cite{Dakic}, $\cH_B$ and $\cH_C$ is not entangled and even have
vanishing discord by the local measurements on the system $\cH_C$.

In fact, let \bea \ptr{A}{\out{u_i}{u_j}} = 0 \quad \textrm{for any
two}\ i\neq j\ \textrm{and}\ p_i,p_j>0, \eea \bea \ket{\Psi}_{ABC} =
\sum_{i\in\Sigma}\sqrt{p_{i}}\ket{u_i}\ket{v_i} \eea be the Schmidt
decomposition of $\ket{\Psi}_{ABC}$. It follows from the condition
that for any $i$, we have \bea \ket{u_i} =
\sum_{j\in\Sigma_i}\sqrt{q^{(i)}_{j}}\ket{\phi^{(i)}_j}\ket{\varphi^{(i)}_j},
\eea where $\{\ket{\phi^{(i)}_j}\}$ and $\{\ket{\varphi^{(i)}_j}\}$
are orthonomal families of $\cH_A$ and $\cH_B$ respectively, and
$\iinner{\phi^{(i)}_j}{\phi^{(k)}_l} = 0$ for $i,  k\in\Sigma, i\neq
k$ and $j\in\Sigma_i,\ l\in\Sigma_k$. Thus, \bea
\rho_{BC} &=& \ptr{A}{\out{\Psi}{\Psi}_{ABC}}\\
&=& \sum_{k\in\Sigma}\sum_{l\in\Sigma_k}\bra{\phi^{(k)}_l}(\sum_{i\in\Sigma}\sum_{j\in\Sigma_i}\sqrt{p_{i}q^{(i)}_{j}}\ket{\phi^{(i)}_j}\ket{\varphi^{(i)}_j}\ket{v_i})
(\sum_{i\in\Sigma}\sum_{j\in\Sigma_i}\sqrt{p_{i}q^{(i)}_{j}}\ket{\phi^{(i)}_j}\ket{\varphi^{(i)}_j}\ket{v_i})^\ast\ket{\phi^{(k)}_l}\\
&=&\sum_{i\in\Sigma}p_{i}\Big(\sum_{j\in\Sigma_i}q^{(i)}_{j}\out{\varphi^{(i)}_j}{\varphi^{(i)}_j}\Big)\ot\out{v_i}{v_i}.
\eea Therefore, $\rho_{BC}$ is a separable state, and note that
$\iinner{v_i}{v_k} = 0$ for all $i\neq k \in \Sigma$, we have
$\cD_{C}(\rho_{BC}) = 0$. The conclusion is proved.

\subsection*{Acknowledgement}  This  project is supported by National Natural Science Foundation of China (11171301, 11571307)
and by the Doctoral Programs Foundation of the Ministry of Education of China (J20130061).





\begin{thebibliography}{99}

\bibitem{Davies}
B. E.~Davies,
\newblock {\it Quantum Theory of Open Sistems}.
\newblock {New York : Academic (1976).}

\bibitem{Holevo}
A. S.~Holevo,
\newblock {\it Probabilistic and Statistical Aspects of Quantum Theory}.
\newblock {Amsterdam : North-Holland (1982).}

\bibitem{Kraus}
K.~Kraus,
\newblock {\it States, Effects and Operations : Fundamental Notions of Quantum Theory}.
\newblock {Berlin : Springer (1983).}

\bibitem{Spehner}
D. Spehner,
\newblock {\it Quantum Correlations and Distinguishability of Quantum States}.
\newblock {J. Math. Phys. \textbf{55}, 075211 (2014).}

\bibitem{Watrous}
J. Watrous,
\newblock {\it Theory of Quantum Information}.
\newblock {Institute for Quantum Computing, University of Waterloo (2008).}

\bibitem{Shannon}
C. E. Shannon,
\newblock {\it A Mathematical Theory of Communication}.
\newblock {Bell Syst. Tech. J. \textbf{27}, 379-423 and 623-656 (1948).}



\bibitem{Ollivier}
H. Ollivier and W. H. Zurek,
\newblock {\it Quantum Discord: A Measure of the Quantumness of Correlations}.
\newblock {Phys. Rev. Lett. \textbf{88}, 017901 (2001).}

\bibitem{Henderson}
L. Henderson and V. Vedral,
\newblock {\it Classical, Quantum and Total Correlations}.
\newblock {J. Phys. A\textbf{34}, 6899-6905 (2001).}

\bibitem{Vedral}
V. Vedral,
\newblock {\it Classical Correlations and Entanglement in Quantum Measurements}.
\newblock {Phys. Rev. Lett. \textbf{90}, 050401 (2003).}


\bibitem{Bennett}
C. H. Bennett, D. P. Di Vincenzo, J. A. Smolin and W. K. Wootters,
\newblock {\it Mixed State Entanglement and Quantum Error Correction}.
\newblock {Phys. Rev. A\textbf{54}, 3824 (1996).}

\bibitem{Wootters}
W. K. Wootters,
\newblock {\it Entanglement of Formation of an Arbitrary State of Two Qubits}.
\newblock {Phys. Rev. Lett. \textbf{80}, 2245 (1998).}

\bibitem{Horodecki}
P.~Horodecki, R.~Horodecki and M.~Horodecki.
\newblock {\it Entanglement and thermodynamical analogies}.
\newblock {Acta Phys. Slov. \textbf{48}, 141 (1998).}

\bibitem{Dakic}
B. Dak\'{\i}c, V. Vedral, and $\check{\textrm{C}}$. Brukner,
\newblock {\it Necessary and Sufficient Condition for Nonzero Quantum Discord}.
\newblock {Phys. Rev. Lett. \textbf{105}, 190502 (2010).}




\end{thebibliography}
\end{document}